\documentclass[runningheads]{svmult}
\usepackage{makeidx}   
\usepackage{graphicx}  
\usepackage{subeqnar}  
\usepackage{multicol}  
\usepackage{cropmark} 
\usepackage{physprbb}  
\begin{document}
\title*{Cosmology with a long-range repulsive force}
\toctitle{Cosmology with a long-range repulsive force}
%
%
\titlerunning{Cosmology with a long-range repulsive force}
%
\author{William H. Kinney\inst{1}
\and Martina Brisudova\inst{2}
\and Richard Woodard\inst{3}}
\authorrunning{W. H. Kinney et al.}
%
%
\institute{Institute for Strings, Cosmology and Astroparticle Physics\\ 
Columbia University\\
Mailcode 5247\\ 
550 W. 120th St. New York, NY 10027
\and Nuclear Theory Center\\
Indiana University\\
2401 Milo B. Sampson Lane\\ 
Bloomington, IN 47408
\and Department of Physics\\ 
University of Florida\\ 
P.O. Box 118440\\ 
Gainesville, FL 32611}

\maketitle              

\begin{abstract}
We discuss the properties of a cosmology dominated by a charged scalar field with a repulsive, long-range self-interaction. The interaction, in the form of a vector field with a tiny mass, can have a dramatic effect on the evolution of the universe, with interesting consequences -- including in some cases accelerated expansion. One characteristic of the model is an oscillating deceleration parameter, which would potentially allow it to be distinguished from other scalar field models such as quintessence.
\end{abstract}

\section{Introduction}

Recent observations of Type Ia supernovae with high redshift indicate that the
universe is entering a phase of cosmological acceleration \cite{reiss,perlm}.
Identifying the causative agent is perhaps the most exciting task for
fundamental theory at present. There are many candidates. It could be a 
cosmological constant, the need for which
was suggested on the basis of other evidence even before the supernovae
results \cite{KT,OS,LLVW}.  Minimally coupled scalars becoming
dominant at late times was also suggested before the supernovae results
\cite{PR,Wett,WCL,FJ}. Since then such models have been dubbed 
``quintessence'' \cite{CDS} and have received extensive study
\cite{CLW,ZWS,BM,BCN,Arm}. Nonminimal couplings have also been explored
\cite{APMS} and recent inspiration has been derived from string theory 
\cite{FKPMP,HKS} and extra dimensions \cite{Maeda,HL}. It has even been
suggested that quantum effects may be responsible \cite{LPR}.

While there is no shortage of models for the physics behind the 
cosmological expansion, there is no single model which can be considered 
compelling.  We have long advocated that 
there might be interesting cosmological implications from long-range forces 
other than classical gravity \cite{TW,us1}. In this paper, we investigate
the cosmological implications of a universal, {\em repulsive} long-range force\footnote{For a more detailed discussion and additional references, see Ref. \cite{us2}.}. 
We wish to evaluate
the simplest scenario which contains the physics we wish to study: a
uniform cosmological scalar with a nonzero local $U(1)$ charge density 
(scalar QED). We find that (perhaps counterintuitively) such universal
self-repulsion does not in itself create accelerated expansion. However,
in combination with a suitably chosen potential for the scalar, an 
accelerating expansion with novel observational signatures can be obtained.
In particular, the model predicts an {\em oscillating} deceleration parameter,
which could be detected by a sufficiently sensitive observation such as
SNAP\cite{SNAP}.

\section{Vector long range force}
\label{seclongrangeforce}

A simple model for a repulsive cosmological force is a complex scalar field 
coupled to a $U(1)$ gauge field. For example, one might choose a Lagrangian of 
the form
\begin{equation}
{\cal L}_0 = \sqrt{-g} \left[ g^{\mu \nu} \left(D_\mu \phi\right)^* \left(D_\nu \phi\right)
 - {1 \over 4} g^{\alpha \rho} g^{\beta \sigma} F_{\alpha \beta} 
F_{\rho \sigma} - V\left(\left\vert \phi \right\vert\right)\right]
\; .
\end{equation}
In a homogeneous fluid with a net charge density, the fluid will be 
self-repulsive. However, realizing this situation in an homogeneous, isotropic 
cosmology is problematic. For example, on a closed 3-manifold, the total charge
of any infinite-range force field must be zero, so the charge density must 
vanish. (This is a consequence of the inability to define the ``inside'' and 
``outside'' of an arbitrary Gaussian surface.) On an open manifold, it is 
possible to impose a nonzero charge density, but only at the expense of 
isotropy. One must choose a boundary condition at infinity which selects a 
direction for the lines of force. These obstacles can be evaded by simply 
making the vector massive \cite{paper1}, for example with an explicit Proca 
term:
\begin{equation}
{\cal L} = {\cal L}_0 + {1 \over 2} m^2 g^{\mu \nu} A_\mu A_\nu \sqrt{-g} \; .
\label{eqbrokenlagrangian}
\end{equation}
The current density $J_\mu \equiv i e \left[\phi \left(D_\mu \phi\right)^* - \phi^*\left(D_\mu \phi\right)\right]$
is still conserved as a consequence of global $U(1)$ invariance. The unique solution consistent with homogeneity and isotropy is
\begin{equation}
A_0 = {i e \left(\phi \partial_0 \phi^* - \phi^* \partial_0 \phi\right) \over 
m^2 + 2 e^2 \phi^* \phi} \; . \label{eqA0solution}
\end{equation} 
The spatial components of the vector potential vanish.

The question we wish to ask is: how does a spacetime dominated by such a 
charged scalar evolve?  The stress-energy tensor is defined by $T_{\mu \nu} \sqrt{-g} \equiv 2 {\delta  S}/{\delta g^{\mu \nu}}$ and its nonzero components in this geometry are given by the pressure and energy density,
\begin{eqnarray}
\rho & = & \left(D_0 \phi\right)^* \left(D_0 \phi\right) + {1 \over 2} 
m^2 A_0^2 + V\left(\phi\right) \; , \\
p & = & \left(D_0 \phi\right)^* \left(D_0 \phi\right) + {1 \over 2} m^2 
A_0^2 - V\left(\phi\right) \; .
\end{eqnarray}
We see immediately that the repulsive interaction contributes a term which 
obeys $p = \rho$, rather than the $p < - \rho/3$ needed for acceleration. This
implies that the new term redshifts very rapidly, as $a^{-6}$, and 
quickly becomes negligible. Physically this is because the Universe expands 
while the mass remains constant, so the force eventually becomes short-range
on cosmological scales. Therefore, the mere presence of a repulsive force does 
not generically lead to acceleration.

However, a fully gauge invariant theory will not have a constant mass for the vector particle: the mass of the vector will be determined by some scalar field via symmetry breaking. The simplest choice is to make the symmetry breaking field the charged field $\phi$, such that
$m^2 \rightarrow 2 \lambda^2 \phi^2$.
The mass of the vector field vanishes at $\phi = 0$, and the force is infinite range. The self-interaction then diverges in the massless limit, and we expect
interesting cosmological effects if 
we begin with a nonzero charge density and then drive the vector mass towards 
zero. It is convenient to decompose the scalar into a magnitude and a phase,
\begin{equation}
\phi\left(t\right) \equiv f\left(t\right) e^{i \theta\left(t\right)} \; ,
\end{equation}
so that $A_0$ depends only upon the phase, $A_0 \propto \dot \theta$.
The energy and pressure depend only on the field magnitude $f$:
\begin{eqnarray}
\rho & = & \dot f^2 + {K \over a^6 f^2} + V\left(f\right) \; , \\
p & = & \dot f^2 + {K \over a^6 f^2} - V\left(f\right) \; . \label{rhop}
\end{eqnarray}
Note that the interaction {\em increases} the energy density, as one would 
expect for a repulsive force. However, the new term in the 
energy density and pressure still looks like a fluid component with 
equation of state
$p = \rho$, and redshifts like $1 / a^6$. One might therefore expect that the 
interaction becomes negligible at late times. This is indeed true when the 
potential minimum occurs at some nonzero value of $f$. However, the situation 
is more interesting when the minimum is at $f = 0$, because then the scalar is
prevented from rolling down to its minimum by the electromagnetic interaction.
For simplicity, we will assume a potential with a monomial form:
\begin{equation}
V\left(f\right) = V_0 f^b .
\end{equation}
The full set of equations describing the dynamics of the system is
\begin{equation}
\ddot f + 3 \left({\dot a \over a}\right) \dot f - {K \over a^6 f^3} + 
{1 \over 2} b V_0 f^{b - 1} = 0,\label{eqfieldEOM0}
\end{equation}
and
\begin{equation}
\left({\dot a \over a}\right)^2 = {8 \pi \over 3 m_{\rm Pl}^2} \left[\dot 
f^2 + {K \over a^6 f^2} + V_0 f^b\right] \; . \label{eqfullEOM0}
\end{equation}
The constant $K$ is related to the charge density $J^0$ by:
\begin{equation}
K \equiv \left({\lambda^2 +  e^2 \over 4 e^2 \lambda^2}\right) \left(a^3 
J^0\right)^2 = {\rm const.}
\end{equation}
These equations are those of a field moving in an 
effective potential which depends upon the scale factor as well as the scalar
magnitude,
\begin{equation}
V_{\rm eff}\left(f,a\right) \equiv V(f) + {K \over a^6 f^2} \; ,
\end{equation}
and we see that the presence of the interaction gives rise to an ``electromagnetic barrier'' which prevents the scalar field from relaxing to the origin. 
(This is similar to the potential proposed for scalar matter with a {\em global} $U(1)$ charge, or ``spintessence'' \cite{GH,BCK}.) Due to the dependence upon $a$ the minimum of the potential $f_0$ is changing as the Universe expands,
\begin{equation}
f_0 \propto  a^{-6 / \left(b + 2\right)} \; . \label{eqadiabaticminimum}
\end{equation}
We can use this as a basis for a self-consistent solution to the equations of motion, such that the field adiabatically follows the minimum of the potential:
\begin{eqnarray}
f(t) &&\propto f_0 \propto  a^{-6 / \left(b + 2\right)},\cr
H(t) &&\propto  \ a^{-3 b/ \left(b + 2\right)}, \label{H0}
\end{eqnarray}
The solution such that $a(t = 0) = 0$ is
\begin{equation}
a(t)\propto t^{\left(b+ 2\right) / 3 b}.\label{eqmasterscalefactor}
\end{equation}
In this case the effective equation of state of the scalar field is determined entirely by the exponent $b$. For example, $b=2$ corresponds to pressureless dust; $b=4$ is radiation, $p = (1/3) \rho$. However, of particular interest are scalar potentials that can provide an accelerated expansion, $p \leq -(1/3) \rho$, which corresponds to $b<1$.

For arbitrary initial conditions, however, the field does not smoothly follow the minimum $f_0$ of the effective potential. The general solution is a field which oscillates about the minimum, with frequency
\begin{equation}
\omega^2 \propto a^{6 \left(2 - b\right) / \left(2 + b\right)}.\label{eqlatetimefreq}
\end{equation}
The oscillations are driven by the time-dependence of the minimum $f_0$, and do not damp with time. Despite the oscillation, however, the behavior ({\ref{eqmasterscalefactor}) of the scale factor is approximately maintained. Figure 1 shows the evolution of the field as a function of scale factor.
\begin{figure}
\begin{center}
\includegraphics[width=.5\textwidth]{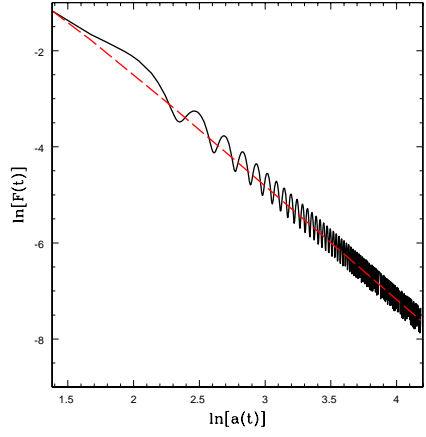}
\end{center} 
\caption{Field $F \equiv \sqrt{8 \pi / 3 m_{\rm Pl}^2} f$ vs. time for $b = 1/2$. The dashed line is the solution (\ref{eqadiabaticminimum}) for $f_0$. }
\end{figure}
Figure 2 shows the three components of the stress-energy (potential, kinetic, and barrier) as the field evolves. Although the potential dominates throughout the evolution, all three terms scale identically with $a$. The precise relationship between the three components is very sensitive to initial conditions. Note in particular the fact that the kinetic energy scales identically with the other terms in the stress-energy, unlike in the case of the self-consistent solution $f_0$.
\begin{figure}
\begin{center}
\includegraphics[width=.5\textwidth]{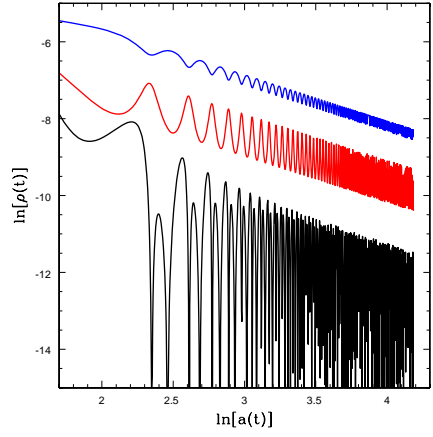}
\end{center} 
 \caption{The three components of the energy density vs. scale factor for $b = 
 1/2$. The potential (upper line, blue) and ''barrier'' (middle line, red) terms dominate the energy density, while the kinetic term (lower line, black) is subdominant. All three scale identically with $a$.}
 \end{figure}
Of most interest are, of course, the cosmological parameters. Do the small 
oscillations of the field affect the background? The deceleration parameter $q$ follows the behavior of the field (Fig. 3). Note in particular that despite the fact that the kinetic energy of the field vanishes when $q$ is at its minimum, the equation of state never reaches $p = -\rho$ due to the contribution of the barrier term to the stress-energy, and $q > -1$ at all times. The system, however is on average in a state of accelerated expansion.
\begin{figure}
\begin{center}
\includegraphics[width=.5\textwidth]{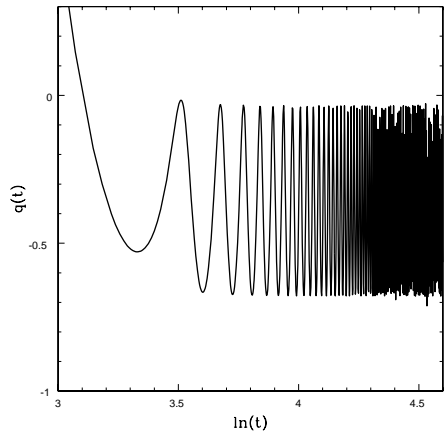}
\end{center} 
\caption{The deceleration parameter $q$ vs. time. The time variable is in the internal units of the numerical solution, $t = [20, 100]$}
\end{figure}
Like the deceleration parameter, the Hubble parameter $H$ also exhibits oscillations. Although the specifics of the field evolution are model dependent, they arise from the field oscillating about the minimum of the effective potential created by the potential and barrier terms. So, oscillations can be expected in any model of this type.  The magnitude of the oscillations on the Hubble diagram can be shown by plotting the residual in the distance modulus $\Delta(m - M)$ between the exact solution and the self-consistent solution, shown in Fig. 4. Figure 5 shows a plot of the deceleration parameter $q$ as a function of redshift for the same choice of model parameters.
\begin{figure}
\begin{center}
\includegraphics[width=.5\textwidth]{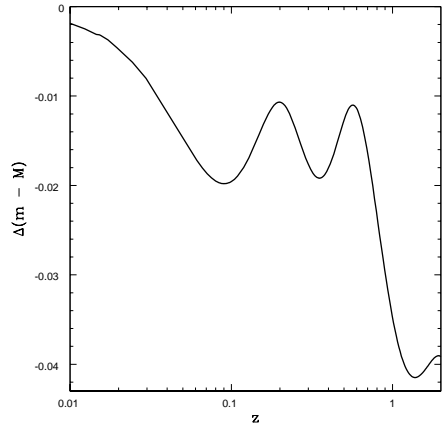}
\end{center} 
\caption{Residual $\Delta(m - M)$ vs. redshift, relative to the self-consistent solution. The oscillations are visible at the level of $\Delta(m - M) \sim 0.04$.}
\end{figure}
\begin{figure}
\begin{center}
\includegraphics[width=.5\textwidth]{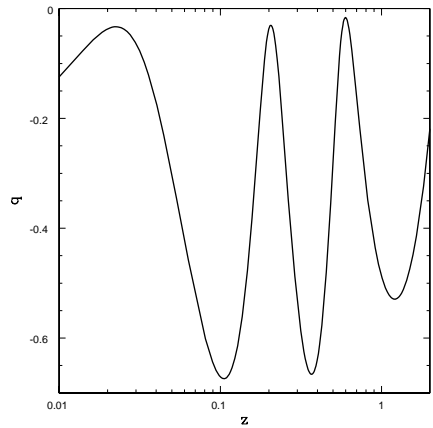}
\end{center} 
\caption{Deceleration parameter $q$ vs. redshift, for reference to Figs. 6 and 7.}
\end{figure}
The magnitude of the oscillations is $\Delta(m - M) \sim 0.04$,  smaller than the accuracy of current supernova Ia observations, which have errors on the order of $\Delta(m - M) \sim 0.15\ -\ 0.20$\cite{SNIa}. Other choices of initial conditions result in a similar order of magnitude for $\Delta(m - M)$, to within a factor of a few. Detection of such a signature would present a formidable observational challenge, but it is conceivable that future measurements such as those by SNAP\cite{SNAP} could approach this level of accuracy. We note with interest that there is in fact some evidence of small scatter beyond observational uncertainties in existing data, at the level of .12 magnitude\cite{SNIa}.

\section{Summary and Conclusion}

The expansion of the universe is observed to be accelerating. However, what physics is responsible for the expansion is at present a completely open question. In this paper we look at a cosmological model dominated by a charged scalar with a long-range gauge interaction, i.e. a repulsive force. The simplest such model, one with an explicitly broken $U(1)$ gauge interaction, is sufficient to show that a universal repulsive force does not in general lead to accelerated expansion. However, the presence of a long-range interaction can substantially alter the dynamics of a scalar field. Cosmological acceleration is possible in a model in which the mass of the vector field vanishes at the minimum of the scalar potential, creating an ``electromagnetic barrier'' that prevents the field from dynamically relaxing to its minimum. Such a model is distinguished by the fact that it predicts an oscillating equation of state, which could be observed by a sufficiently precise observation of the distance modulus/redshift relation for distant supernovae, such as that envisioned for SNAP\cite{SNAP}.

\section*{Acknowledgments}
This work was partially supported by DOE contract DE-FG02-97ER-41029 and by
the Institute for Fundamental Theory. WHK is supported by the Columbia 
University Academic Quality Fund. ISCAP gratefully acknowledges the generous
support of the Ohrstrom Foundation.

\end{document}